\documentclass[epj]{svjour}
\usepackage{graphics}
\usepackage{bm}   
\usepackage{color}
\usepackage{amssymb}
\usepackage{amsmath}
\usepackage{balance}

\newcommand{\sech}{\normalfont\mbox{sech}\,}
\newcommand{\e}{\normalfont\mbox{e}\,}

\begin{document}

\title{Amplification of matter rogue waves and breathers in quasi-two-dimensional Bose-Einstein condensates}
\author{K. Manikandan \inst{1} M. Senthilvelan \inst{1,} \thanks{\emph{e-mail:velan@cnld.bdu.ac.in}} \and R.A. Kraenkel \inst{2}}

\institute{Centre for Nonlinear Dynamics, Bharathidasan University, Tiruchirappalli 620024, Tamilnadu, India \and Instituto de F\'{\i}sica Te\'orica , Universidade Estadual Paulista, Rua Dr. Bento Teobaldo Ferraz 271, 01140-070 S\~ao Paulo, Brazil}

\abstract{
We construct rogue wave and breather solutions of a quasi-two-dimensional Gross-Pitaevskii equation with a time-dependent interatomic interaction and external trap.  We show that the trapping potential and an arbitrary functional parameter that present in the similarity transformation should satisfy a constraint for the considered equation to be integrable and yield the desired solutions.  We consider two different forms of functional parameters and investigate how the density of the rogue wave and breather profiles vary with respect to these functional parameters.  We also construct vector localized solutions of a two coupled quasi-two-dimensional Bose-Einstein condensate system.  We then investigate how the vector localized density profiles modify in the constant density background with respect to the functional parameters. Our results may help to manipulate matter rogue waves experimentally in the two-dimensional Bose-Einstein condensate systems.
} 
\titlerunning{Rogue waves and breathers in two-dimensional Bose-Einstein condensates}
\authorrunning{K. Manikandan et al.}
\maketitle

\section{Introduction}
During the past two decades, experimental achievements and theoretical studies on Bose-Einstein condensates (BEC) of weakly interacting atoms have stimulated a broad interest in the field of atomic matter waves and nonlinear excitations \cite{dark,bright,gap,vort,blud:kono}.  At absolute zero temperature, the properties of a condensates can be described by the mean-field Gross-Pitaevskii equation (GPE) (which is a generalized form of the ubiquitous constant coefficient nonlinear Schr\"{o}dinger (NLS) equation) with suitable nonlinearity \cite{peth:smi,pit:str}.  For example, the cubic and/or quintic nonlinearity in the GP equation represents the two-body interaction and/or three-body interaction respectively.  The three-body interaction is usually neglected to conserve the energy and the total number of atoms. It has been shown that the properties of BEC are determined by the sign and magnitude of the $s$-wave scattering length which can be controlled by means of the external magnetic field or low loss optical Fesbach resonance (FR) technique \cite{dalf,stali,abdul,magn,opti}.  This freedom offers us to consider the coefficient of nonlinear term in the GP equation can be a function of time and/or space.  By utilizing this time/space dependent coefficient, we can manage/control various localized structures such as solitons, breathers and rogue waves (RWs) \cite{atre:pani,serki,raj:mur,yan,wen:li,he,loomba,Mani} profitably.  

Several works have been devoted to study the localized structures in the quasi-one-dimensional BECs \cite{raj:mur,yan,wen:li,he,loomba,Mani}. It is evident from the literature that studies have also been made on the quasi-one-dimensional two component GPEs \cite{raje,kono,vina,babu}.  Eventhough the one-dimensional equations have given a good understanding on the dynamics, a detailed investigation on the higher-dimensional version (two or three dimensions) of the system will provide more clear visualization about the localized structures.  As far as the two-dimensional NLS equations with variable coefficients are concerned only few studies have been made to identify the localized structures \cite{cqdai,zyan,dswang,zjfang,zhu,zyan2}.  The RW and breather profiles of two-dimensional coupled equations have been constructed by mapping them into the scalar NLS equation \cite{dai2,kkde}. Very recently an attempt has been made to transform the two-dimensional two coupled variable coefficients NLS equation into two coupled constant coefficient NLS equation \cite{wang3}. However, the results which we report here are more general than the one reported in \cite{wang3}.

In this paper we consider both single and two component GPEs in two-dimensions and investigate the dynamical evolutions of matter RWs and breathers in them.  RW is a nonlinear wave which appears from nowhere and disappears without a trace \cite{osbrn:rato}. It has a wave-height (the distance from trough to crest) which is two or three times greater than the significant wave height \cite{osbrn,khar:pelin}.  RWs arise due to the instability of a certain class of initial conditions that tend to grow exponentially and hence have the possibility of increasing up to very high amplitudes \cite{benj:feir}.  The RWs have also been observed experimentally in several physical systems including water wave tank \cite{chab}, capillary waves \cite{shatz} and nonlinear optics \cite{solli,kibler}.  Efforts also have been made to find the explicit expressions for the RWs. One of the rational solutions of the constant coefficient NLS equation, the Peregrine soliton \cite{pere}, a localized solution in both space and time, can be used to model the RW.  Several theoretical studies on the dynamics of RWs in nonlinear fiber optics \cite{solli}, plasma physics \cite{mose}, laser-plasma interactions \cite{veldas}, and even econophysics \cite{tan} have been carried out in recent times. The higher-order rational solutions of the NLS equation (represent the higher order RWs) were constructed in \cite{akmv:anki}. A few studies have also been devoted to investigate RWs in vector NLS equations.  In this direction exact analytical RW solutions for the completely integrable Manakov equation \cite{ling,fabio,vishnu,lomb} and its generalizations were constructed in \cite{vishnu3}. 

Physically breathers arise from the effect of modulational instability which is a characteristic feature of various nonlinear dispersive systems and associated with dynamical growth and evolution of periodic perturbation on a continuous background \cite{mande,eleon}. Breathers have been categorized into two main kinds: (i) Akhmediev breather (AB) (periodic in space and localized in time) and (ii) Ma breather (MB) (periodic in time and localized in space) \cite{eleon}.

In the context of BEC, the formative mechanism of the matter RWs is the accumulation of energy and atoms towards its central part and their spreading out to a constant density background.  The formation of matter in breather form in BEC corresponds to the periodic exchange of atoms between the profile and the plane wave background.  In experiments, RW and breather structures in BEC systems can be created and controlled by tuning of nonlinear interaction between atoms through Feshbach resonance technique \cite{dalf,stali,abdul} and by modulating the trapping frequency of the external potential.  Motivated by this experimental feasibility, we intend to explore the dynamics of RWs and breathers in two-dimensional BECs.  We consider two-dimensional single and two-component GPE with a variable nonlinearity parameter and external trap potential and contruct several localized solutions of them. To construct these solutions we map the two-dimensional time-dependent GP equation into constant coefficient NLS equation through similarity transformation. We show that the trapping potential and an arbitrary functional parameter that present in the similarity transformation should satisfy a constraint for the considered equation to be integrable and yield the desired solutions. From the known RW and breather solutions of the constant coefficient NLS equation we present RW and breather solutions of the two-dimensional GP equation.  We consider two different forms of functional parameters and investigate how the density of RW and breather profiles amplify with respect to these functional parameters.  Our results show that in the case of RW, its amplitude increases and become more localized in time whereas in the case of breather, the number of peaks increases in the underlying profiles.  In addition to the above, we construct vector localized solutions for the two coupled quasi-two-dimensional BECs.  Here we investigate how the vector localized density profiles amplify in a constant density background when we vary the functional parameter.  Our results reveal that the density profiles become more localized and their amplitudes become sharpen in the density background when we vary the first functional parameter whereas in the case of second functional parameter the underlying structures become more localized in time and delocalized in space.

We organize our work as follows.  In Sec. 2, we consider the model for quasi-two-dimensional BECs and map the time-dependent two-dimensional GP equation into constant coefficient NLS equation using similarity reduction technique.  In Sec. 3, we construct RW solutions without and with free parameters and analyze how the nature of fundamental pattern of the RWs amplify in the density background by varying the arbitrary functional parameter.  Further, we construct one-breather and two-breather solutions of the two-dimensional GP equation and investigate their characteristics with respect to the functional parameter.  In Sec. 4, we construct the vector localized solutions of a two coupled quasi-two-dimensional BECs and examine in detail how their associated characteristics change when we vary the arbitrary functional parameter. Finally, in Sec. 5, we present our conclusions. 

\section{The Model and RW solutions}
In mean-field theory, the BEC at low temperature is described by the three dimensional GP equation.  Let us suppose that the BEC is confined in a harmonic trap $v(x,y,z)=\frac{m}{2}(\omega_{r} r^2+\omega_z z^2)$, where $m$ is atomic mass, $r^2=x^2+y^2$ and $\omega_{r}$, $\omega_z$ are the confinement frequencies in the radial and axial directions, respectively.  If the trap is pancake shaped, that is $\omega_z \geq \omega_{r}$, it is reasonable to reduce the GP equation for the condensate wave function to an quasi-two-dimensional equation \cite{peth:smi,pit:str}, namely
\begin{equation}
i\psi_t+\frac{f}{2}\left(\psi_{xx}+\psi_{yy}\right)+R(t)\vert \psi \vert^2 \psi+ \frac{1}{2}\beta^2(t)(x^2 +y^2) \psi=0,
\label{2d:eq1}
\end{equation}
where $\psi(x,y,t)$ denotes the macroscopic condensate wave function, $t$ is the temporal and $x$ and $y$ are the spatial coordinates measured in units of $\omega_{r}^{-1}$ and $a_{r}=\sqrt{\hbar/(m\omega_{r})}$, respectively and $f$ is the dispersion constant.  The nonlinearity parameter is defined by $R(t)=\frac{2 a_s(t)}{a_B}$, where $a_s(t)$ is the $s$-wave scattering length and $a_B$ is the Bohr radius. The trapping potential parameter is defined by the expression $\beta^2(t)=\frac{\omega_{r}^2(t)}{\omega_z^2}$.  The functions $R(t)$ and $\beta(t)$ can be controlled experimentally.  We confine our attention only on attractive interatomic interactions.

To investigate the dynamics of (\ref{2d:eq1}), we map the two-dimensional GP equation $(\ref{2d:eq1})$ to standard NLS equation through the similarity transformation
\begin{equation}
\psi(x,y,t)=r(t)U(X,T)\exp[i \theta(x,y,t)],
\label{2d:eq2}
\end{equation}
where $r(t)$ is the amplitude, $T(t)$ is the effective dimensionless quantity, $X(x,y,t)$ is the similarity variable and $\theta(x,y,t)$ is the phase factor which are all to be determined.  To determine these unknown functions, we substitute $(\ref{2d:eq2})$ into $(\ref{2d:eq1})$ and obtain the following set of partial differential equations for these unknowns, namely
\begin{eqnarray}
&& X_{xx}+X_{yy}=0, \notag \\
&& X_t+f(X_x\theta_x+X_y\theta_y)=0, \notag \\
&& \frac{r'(t)}{r(t)}+\frac{f}{2}\left(\theta_{xx}+\theta_{yy}\right)=0, \label{pdes}\\ 
&& \theta_t+\frac{f}{2}(\theta_x^2+\theta_y^2)-\frac{1}{2}\beta^2(t)(x^2+y^2)=0,  \notag \\
&& T_t-R(t)r^2(t)=0, \;\;\; f(X_x^2+X_y^2)-R(t)r^2(t)=0. \notag
\end{eqnarray}
Solving the above system of PDEs (\ref{pdes}), we find
\begin{eqnarray}
\label{2d:eq3}
&& k(t) = k_0 \sqrt{F_1'(t)}, \;\;\;\ l(t)= l_0 \sqrt{F_1'(t)}, \;\;\; r(t) = r_0\sqrt{F_1'(t)}, \notag \\
&& X(x,y,t)  = k(t) x + l(t) y+F_1(t), \;\;\; T(t) = \int{R(t)r^2(t)}dt, \notag \\
&& c(t)= -\frac{(k_0^2+l_0^2)F_1(t)}{8f k_0^2l_0^2}, \;\;\; R(t)= \frac{f \left(k^2(t)+l^2(t)\right)}{r^2(t)}, \notag \\
&& \theta(x,y,t)= -\frac{1}{2f}\left(\frac{k'(t)}{k(t)}x^2 + \frac{l'(t)}{l(t)}y^2\right) \notag \\ && \hspace{2cm}-\frac{F_1'(t)}{2f}\left(\frac{x}{k(t)}+\frac{y}{l(t)}\right) +c(t),
\end{eqnarray}
where $F_1(t)$ is an arbitrary functional parameter and $r_0, k_0$ and $l_0$ are arbitrary constants which come out while integrating the system of PDEs (\ref{pdes}). In this process, $R(t)$ is a constant and the function $U(X,T)$ is found to satisfy the standard NLS equation, that is
\begin{equation}
\label{nls}
i \frac{\partial U}{\partial T}+\frac{1}{2}\frac{\partial ^2 U}{\partial X^2}+ |U|^2 U=0.
\end{equation}
The trapping potential and the functional parameter $F_1(t)$ should satisfy the following constraint, namely
\begin{equation}
\label{2d:eq4}
2 \frac{F_1'''(t)}{F_1(t)}-3\left(\frac{F_1''(t)}{F_1'(t)}\right)^2+4f\beta^2(t)=0,
\end{equation}
where prime refers differentiation with respect to $t$.  Irrespective the form of $F_1(t)$, as long as the condition (\ref{2d:eq4}) is satisfied, we can obtain a solution of $(\ref{2d:eq1})$ in the form
\begin{subequations}
\label{2d:eq6}
\begin{align}
\label{2d:eq6a} \psi(x,t)= & r_0\sqrt{F_1'(t)}U(X,T) \eta(x,y,t), \\
\label{2d:eq6b} \eta(x,y,t)= & \exp \left[i\left(-\frac{1}{2f}\left(\frac{k'(t)}{k(t)}x^2 + \frac{l'(t)}{l(t)}y^2\right) \right. \right. \\  & \left. \left. -\frac{F_1'(t)}{2f}\left(\frac{x}{k(t)}+\frac{y}{l(t)}\right) -\frac{(k_0^2+l_0^2)F_1(t)}{8f k_0^2l_0^2}\right)\right], \notag
\end{align}
\end{subequations}
where $U(X,T)$ is the solution of the standard NLS Eq. (\ref{nls}). As we are interested in the RW and breather solutions of (\ref{2d:eq1}) we are going to use only these two solutions of (\ref{nls}).  Eq. (\ref{2d:eq6}) describes the dynamics of localized matter waves in two-dimensional BECs.  
\section{Dynamical evolutions of RWs and breathers}
In this section, we examine the characteristics of RWs and breather solutions of two-dimensional time-dependent GP Eq. (\ref{2d:eq1}) by considering two different forms of functional parameters, namely (i) $F_1(t)=b_2 t^2+b_1 t+b_0 $ and (ii) $F_1(t)= 1+\tanh{(b_0 t)}$, where $b_0, b_1$ and $b_2$ are positive constants.  We will also point out the relevance of these two solutions in the context of condensation of atoms.   
\subsection{Case 1} 
\subsubsection{Characteristics of RWs}
Substituting $F_1(t)=b_2 t^2+b_1 t+b_0 $ in the obtained solution (\ref{2d:eq6}), we find 
\begin{subequations}
\label{2d:a15}
\begin{align}
\label{2d:a15b} \psi(x,y,t) = & r_0\sqrt{2 b_2 t+b_1}\, U_j(X,T)  \eta(x,y,t), \\
\label{2d:a15a} \eta(x,y,t) = & \exp\bigg\{\frac{1}{8f}i\bigg[\frac{4b_2(x^2+y^2)}{b_1+2b_2t}+\frac{4\sqrt{b_1+2b_2t}}{k_0l_0} \\
& \left.\left. \times (k_0 x+l_0 y) +\frac{(k_0^2+l_0^2)(b_0+t(b_1+b_2t))}{k_0^2l_0^2}\right]\right\}, \notag
\end{align}
\end{subequations}
where $U_j(X,T)$'s, $j=1, 2, 3$, are the first, second and third-order RW solutions of NLS Eq. (\ref{nls}) whose explicit expressions are given in the Appendix (see Eqs. (\ref{a8}), (\ref{a11}) and (\ref{a12})). 
\begin{figure*}[!ht]
\begin{center}
\resizebox{0.99\textwidth}{!}{\includegraphics{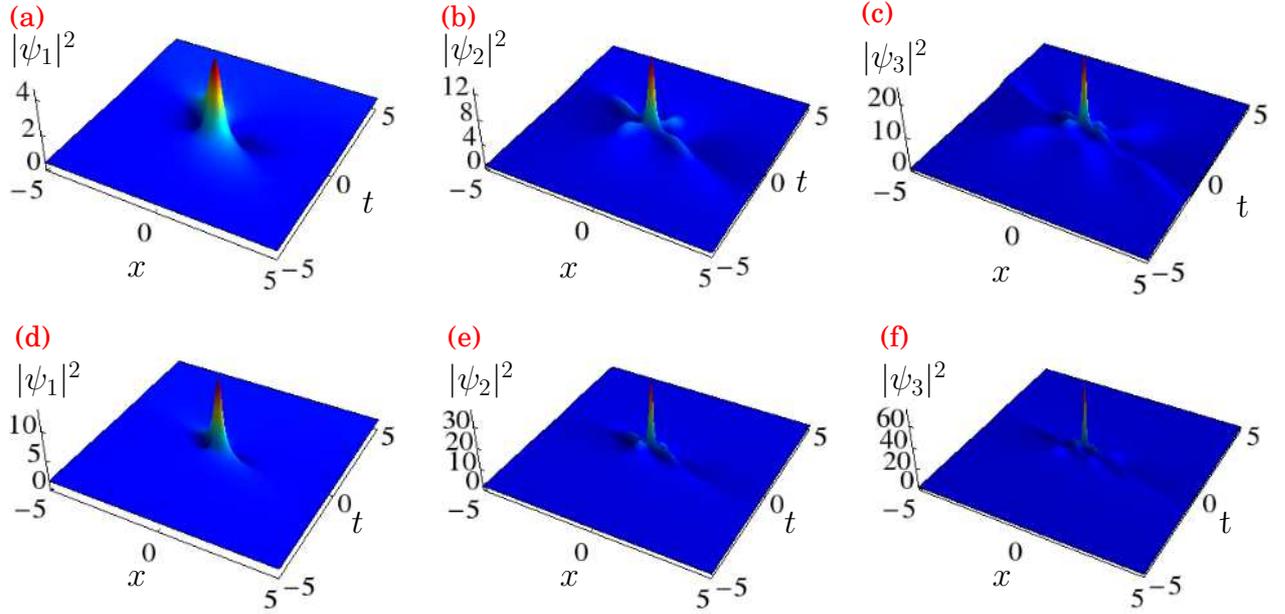}}
\end{center}
\caption{Density profiles.  (a), (d) First-order RW, (b), (e) second-order RW, and (c), (f) third-order RW for $F_1(t)=b_2 t^2+b_1 t+b_0 $.  The parameter $b_1$ is chosen as $0.5$ in (a)-(c) and $1.5$ in (d)-(f).  The other parameters are fixed as $r_0=1.0$, $k_0=l_0=1$, $b_2=0.02$ and $b_0=0.01$.}
\label{2d:fig1} 
\end{figure*}
In Figure \ref{2d:fig1}, we present the localized density profiles of first, second, and third-order RW solutions of Eq. (\ref{2d:eq1}) for the functional parameter $F_1(t)=b_2 t^2+b_1 t+b_0 $. From BEC perspective, the appearance of RWs in (\ref{2d:eq1}) can be identified with the fluctuations in the density of the condensate atoms which are localized in space and time.  For convenience, throughout this work, we choose the parameters $b_2$ and $b_0$ in $F_1(t)$ as $b_2=0.02$ and $b_0=0.01$. The density plot of first-, second-, and third-order RWs with $b_1=0.5$ are presented in Figures~\ref{2d:fig1}(a)-(c). Now we examine how these localized density profiles vary with respect to the arbitrary parameter $b_1$.  When we increase the value of $b_1$ the density profiles become more localized in time and the amplitudes of them become higher as shown in Figures~ \ref{2d:fig1}(d)-(f). 
\begin{figure*}[!ht]
\begin{center}
\resizebox{0.99\textwidth}{!}{\includegraphics{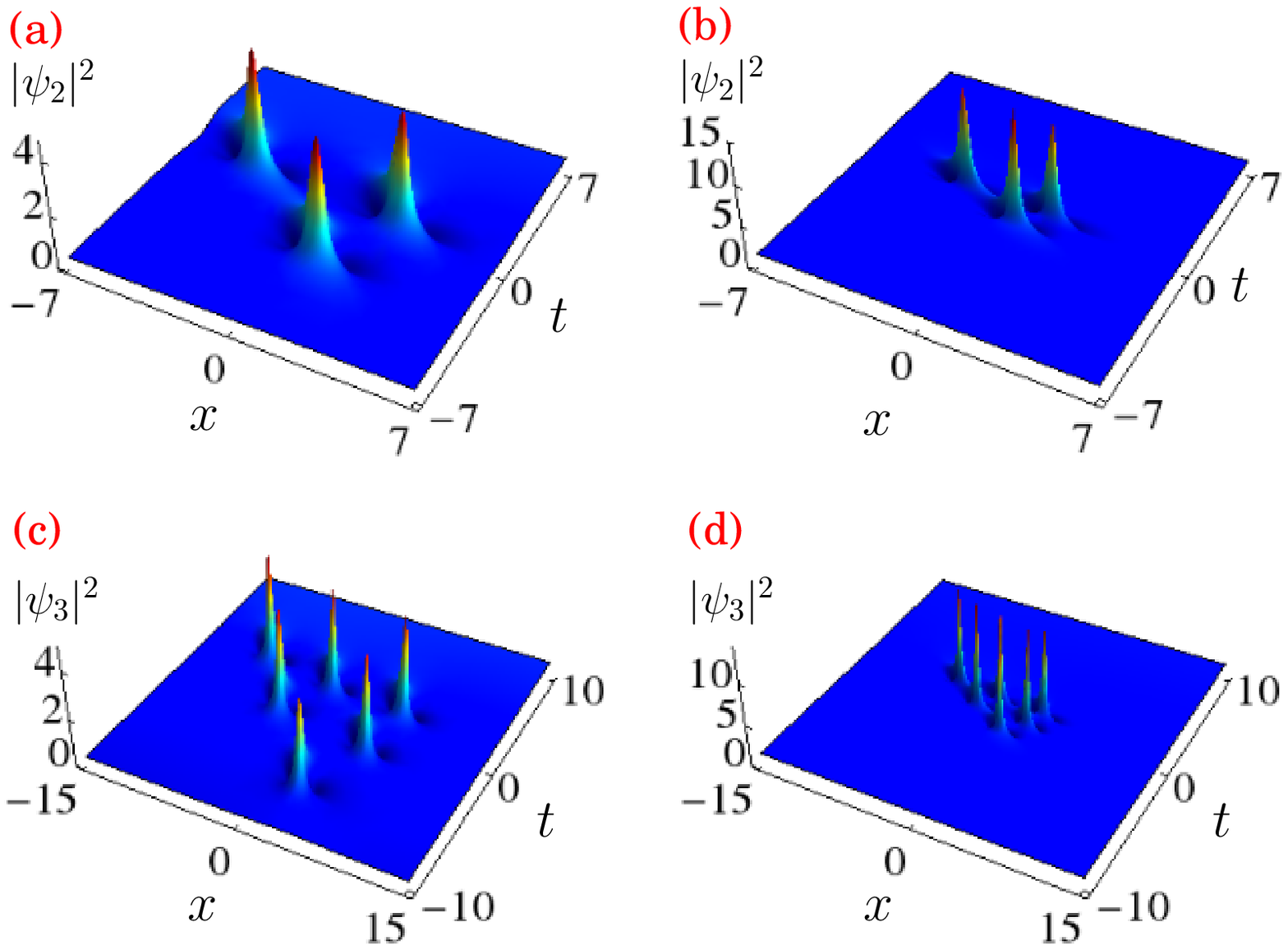}}
\end{center}
\caption{Density profiles.  (a), (b) Triplet RW, and (c), (d) sextet RW for $F_1(t)=b_2 t^2+b_1 t+b_0 $.  The parameter $b_1$ is chosen as $0.5$ in (a),(c) and $1.5$ in (b), (d).  The other parameters are fixed as $r_0=1.0$, $k_0=l_0=1$, $b_2=0.02$ and $b_0=0.01$.}
\label{2d:fig2}
\end{figure*}

One can also generalize the expressions of second and third order RW solutions of the scalar NLS Eq. (\ref{nls}) by introducing certain free parameters in them \cite{ankie}.  These free parameters split the symmetric form of RW solution into a multi-peaked solution and the distances between the peaks depend on these free parameters.  By introducing these free parameters, the higher-order RW solutions decompose into $n(n+1)/2$ first-order forms, where $n$ is the order of the RW \cite{hezhang}. The symmetry of the higher-order solutions are maintained even in the decomposed forms.  By varying these free parameters we can extract certain novel patterns exhibited by the RWs \cite{ankie}. These free parameters determine the size and orientation of the first-order solution as well \cite{ankie}.  Motivated by these recent observations, in the following, we intend to consider the generalized second- and third-order RW solutions of (\ref{2d:eq1}) and investigate how these generalized RW structures modify with respect to the functional parameter $F_1(t)$. 

To begin, we consider the generalized second-order RW solution of (\ref{nls}).  In this case, we have the following modified expressions \cite{ankie} for $G_2$, $H_2$ and $D_2$ in Eq. (\ref{a9}), namely 
\begin{align}
\label{trisoln}
 G_2 = & 12 \left[3-16 X^4-24 X^2(4 T^2+1)-48 s X-80 T^4 \right. \notag \\ & \left. -72 T^2-48 m T\right], \notag \\
 H_2 = & 24 \left\{ T\left[ 15-16 X^4+24 X^2-48 s X-8(1-4 X^2)T^2 \right. \right. \notag \\ & \left. \left. -16 T^4 \right] +6m(1-4 T^2+4X^2)\right\},  \\
D_2 = & 64 X^6+48 X^4 (4T^2+1)+12X^2(3-4T^2)^2+64 T^6 \notag \\ & +432 T^4+396 T^2+9 +48m\left[18m+T(9+4 T^2 \right. \notag \\ & \left.-12X^2)\right] +48 s\big[(18 s+X(3+12T^2 -4X^2)\big]. \notag  
\end{align}
Upon comparing the expressions (\ref{trisoln}) and (\ref{a9}) we can notice that we have two new free parameters, namely $s$ and $m$, are in the generalization. These two parameters describe the relative positions of the first-order RWs in the triplet.  Triplet is symmetry preserving first-order structures revealing the fact that the second-order RW solution is a family of three first-order rational solutions \cite{ankie}.  Substituting (\ref{trisoln}) into (\ref{2d:a15}), for $j=2$, we can capture the generalized second-order RW solution of (\ref{2d:eq1}).  

For the choice $s=m=0$, the expression (\ref{trisoln}) reduces to the fundamental second-order RW solution. With $s\neq m\neq 0$ the generalized second-order RW structure splits into three first-order RWs. These waves emerge in a triangular fashion (a triplet pattern). These three first-order RWs have a structure of equilateral triangle with $120$ degrees of angular separation between them \cite{ankie}.  In Figures \ref{2d:fig2}(a)-(b), we show the triplet pattern for $s=15$ and $m=25$.  The formation of triplet patterns, for the parametric values $b_1=0.5$ and $b_1=1.5$, are shown in Figures~\ref{2d:fig2}(a) and \ref{2d:fig2}(b), respectively.  For $b_1=1.5$ the triplet pattern is more localized in time. The amplitude of it has increased as shown in Figure~\ref{2d:fig2}(b).  

We move on to investigate the structure of the generalized third-order RW solution with the same functional parameter $F_1(t)$. We have four new free parameters, $s_1,s_2,g$ and $h$ in the generalized third-order RW solution.  We do not give the explicit expression of it here since it is very lengthy.  We analyze the results only graphically.  With $s_1= s_2= g= h=0$, we have the fundamental pattern, see Figure~\ref{2d:fig1}(c), with a single maximum at the centre surrounded by six sub-peaks.  For non-zero values of $s_1$, $s_2$, $g$ and $h$, the third-order RW splits into six separated first-order RWs (sextet pattern).  The sextet pattern for the choice $s_1=10$, $s_2=20$, $g=500$ and $h=500$ is demonstrated in Figures~\ref{2d:fig2}(c)-(d).  The sextet pattern for $b_1=0.5$ is depicted in Figure~\ref{2d:fig2}(c).  When we raise the value of this parameter, say for example $b_1=1.5$, the sextet pattern becomes more localized in time.  The density has also increased which can be visualized from Figure~\ref{2d:fig2}(d).  
\subsubsection{Characteristics of breathers}
In the above, we have studied how the localized density profiles of RWs get amplified with respect to the functional parameter $F_1(t)$ for the time-dependent two-dimensional GPE.  We now investigate how the density of breather profiles change in the condensate background when we vary this functional parameter.

To begin, we consider the first-order breather solution of NLS equation (\ref{nls}), namely \cite{eleon}
\begin{align}
\tilde{U_1}(X,T) = & \,\biggr[\frac{q^2 \cosh[\alpha (T - T_1)] + 2 i \,q v \sinh[\alpha (T - T_1)]}{2  \cosh[\alpha (T - T_1)] - 2v \cos[ q (X - X_1) ]} \notag \\
& \, -1\biggr] \times \exp{(iT)},
\label{b1}
\end{align}%
where the parameters $q$ and $v$ are expressed in terms of an arbitrary complex eigenvalue (say $\lambda$), that is $q=2\sqrt{1+\lambda^2}$ and $v = \mbox{Im}(\lambda)$, and $X_1$ and $T_1$ serve as coordinate shifts from the origin.  The real part of the eigenvalue represents the angle that the one-dimensional localized solutions form with the $T$ axis, and the imaginary part characterizes the frequency of periodic modulation \cite{eleon}.  The parameter $\alpha$ $(=q v)$ in (\ref{b1}) represents the growth rate of modulation instability.  Substituting (\ref{b1}) into (\ref{2d:a15}) we can get the breather solution of (\ref{2d:eq1}). Now we study the breather dynamics of (\ref{2d:eq1}).   
\begin{figure*}[!ht]
\begin{center}
\resizebox{0.99\textwidth}{!}{\includegraphics{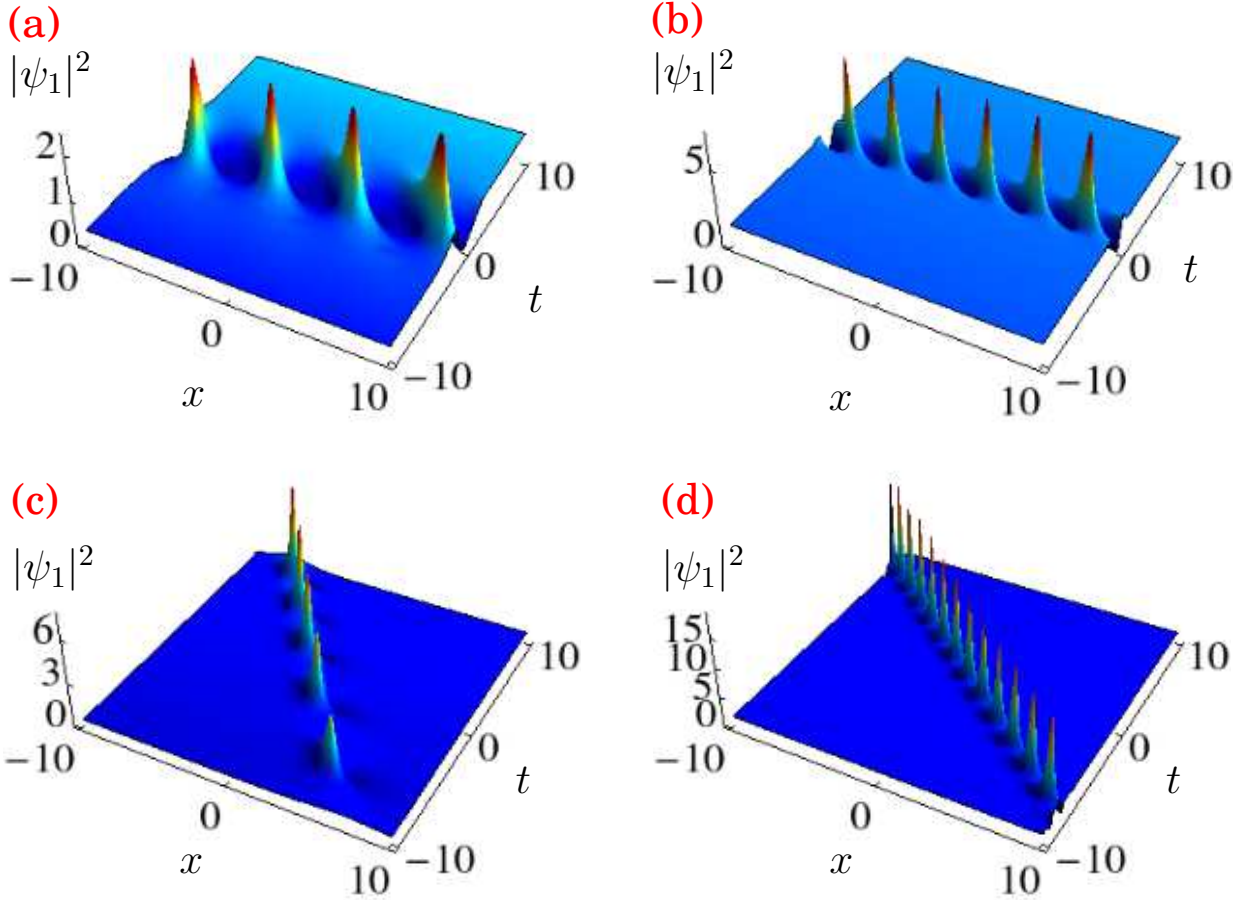}}
\end{center}
\caption{Density profiles.  (a), (b) AB, and (c), (d) Ma breather for $F_1(t)=b_2 t^2+b_1 t+b_0 $.  The parameter $b_1$ is chosen as $0.5$ in (a),(c) and  $1.5$ in (b), (d).  The other parameters are fixed as $r_0=1.0$, $k_0=$, $l_0=1$, $b_2=0.02$ and $b_0=0.01$.}
\label{2d:fig3}
\end{figure*}

The plot of matter breather profiles of (\ref{2d:eq1}) for $F_1(t)=b_2 t^2+b_1 t+b_0$ is depicted in Figure~\ref{2d:fig3}. We can extract AB from (\ref{b1}) for the eigenvalue $\lambda=0.6 i$, which is spatially periodic and localized in time.  The AB is shown in Figure~\ref{2d:fig3}(a). We can deduce an expression for MB, from (\ref{b1}) by choosing the eigenvalue $\lambda=1.4 i$, which is temporally periodic and localized in space.  The MB is shown in Figure~\ref{2d:fig3}(c).  When we increase the value of $b_1$ the number of peaks increases in the constant density background in the considered region $x-t$.  In the case of AB, the number of peaks is 4 when $b_1=0.5$ and it becomes 6 at $b_1=1.5$, in the considered region of $x-t$ plane, which is demonstrated in Figure~\ref{2d:fig3}(b).  In the case of MB, the number of peaks increases from 6 to 12 when we vary the parameter $b_1$ from 0.5 to 1.5 in the considered region of $x-t$ plane which is displayed in Figure~\ref{2d:fig3}(d).  When we increase the value of $b_1$ in the obtained AB and MB solutions, the number of peaks and their density increases. 

In what follows, we will construct the two-breather solution and examine how the underlying structure change with respect to the modulation parameters.  The two-breather solution of NLS equation is given by \cite{kadz}
\begin{align}
\label{2b1}
\tilde{U_2}(X,T)=\left[1+\frac{\tilde{G_2}(X,T)+i \tilde{H_2}(X,T)}{\tilde{D_2}(X,T)}\right]\exp{(iT)},
\end{align}
where $\tilde{G_2}$, $\tilde{H_2}$, and $\tilde{D_2}$ are given by 
\begin{subequations}
\label{2b1a}
\begin{align}
\tilde{G_2} = &\, -(k_1^2-k_2^2) \biggr[\frac{k_1^2\delta_2}{k_2}\cosh(\delta_1T_{s1})\cos(k_2X_{s2}) \notag \\ 
              &\,  - (k_1^2-k_2^2)\cosh(\delta_1T_{s1})\cosh(\delta_2T_{s2}) \notag \\ 
              &\, -\frac{k_2^2\delta_1}{k_1}\cosh(\delta_2T_{s2})\cos(k_1 X_{s1})\biggr],  \\
\tilde{H_2} = &\, -2(k_1^2-k_2^2)\biggr[\frac{\delta_1\delta_2}{k_2}\sinh(\delta_1 T_{s1})\cos(k_2 X_{s2})\notag \\ 
              &\, - \frac{\delta_1 \delta_2}{k_1}\sinh(\delta_2T_{s2}) \cos(k_1 X_{s1}) \notag \\   
              &\,  -\delta_1 \sinh(\delta_1 T_{s1})\cosh(\delta_2 T_{s2}) \notag \\   
              &\,  + \delta_2 \sinh(\delta_2 T_{s2})\cosh(\delta_1 T_{s1})\biggr], \\
\tilde{D_2} = &\,  2 (k_1^2 + k_2^2) \frac{\delta_1 \delta_2}{k_1 k_2} \cos(k_1X_{s1})\cos(k_2 X_{s2}) \notag \\ 
              &\,  + 4 \delta_1 \delta_2 (\sin(k_1 X_{s1}) \sin(k_2 X_{s2})  \notag \\ 
              &\,  + \sinh(\delta_1 T_{s1})\sinh(\delta_2 T_{s2})) \notag \\ 
              &\,- (2 k_1^2 - k_1^2 k_2^2 + 2 k_2^2) \cosh(\delta_1 T_{s1})\cosh(\delta_2 T_{s2}) \notag \\ 
              &\, - 2 (k_1^2 - k_2^2) \biggr[\frac{\delta_1}{k_1}\cos(k_1 X_{s1})\cosh(\delta_2 T_{s2}) \notag \\ 
              &\,  - \frac{\delta_2}{k_2}\cos(k_2 X_{s2}) \cosh(\delta_1 T_{s1})\biggr], 
\end{align}
\end{subequations}
where the modulation frequencies, $k_j=2\sqrt{1+\lambda_j^2}$, $j=1,2$, are described by the (imaginary) eigenvalues $\lambda_j$.  In the above expressions, $X_j$ and $T_j$, $j=1,2$, represent the shifted point of origin, $\delta_j$ $(=k_j\sqrt{4-k_j^2}/2)$ is the instability growth rate of each component and $X_{sj}=X-X_j$ and $T_{sj}=T-T_j$ are the shifted variables.  One can extract a variety of second-order breather structures such as ABs, MBs and the intersection of AB and MB solutions from (\ref{2b1}) for certain combinations of these eigenvalues. For example, we can derive ABs from (\ref{2b1}) when the imaginary parts of both  the eigenvalues, $\mbox{Im}(\lambda_j)$, $j=1,2$, lie between $0$ and $1$. On the other hand the imaginary parts of both the eigenvalues are greater than one ($\mbox{Im}(\lambda_j)>1$) then the solution provides MBs. In the mixed case, that is one of the eigenvalues is less than one and the other is greater than one, we can obtain the intersection of AB and MB solutions.   
\begin{figure*}[!ht]
\begin{center}
\resizebox{0.99\textwidth}{!}{\includegraphics{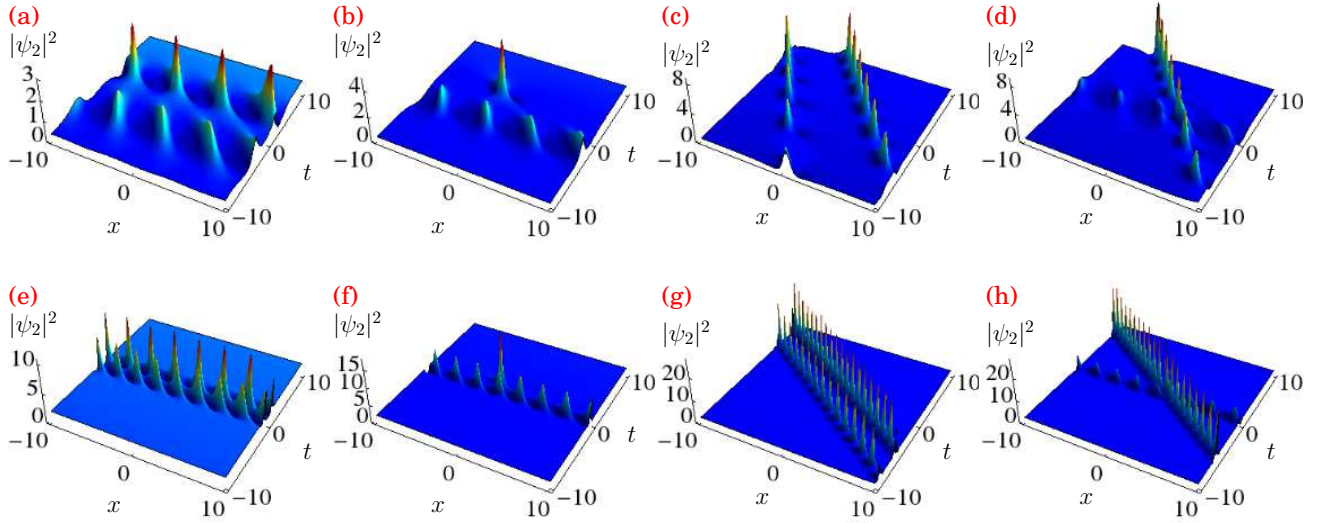}}
\end{center}
\caption{Density profiles.  (a), (e) Two AB, (b), (f) AB - RW, (c), (g) two Ma breather and (d), (h) the intersection of AB-MB for $F_1(t)=b_2 t^2+b_1 t+b_0 $.  The parameter $b_1$ is chosen as $0.5$ in (a)-(d) and $1.5$ in (e)-(h).  The other parameters are fixed as $r_0=1.0$, $k_0=l_0=1$, $b_2=0.02$ and $b_0=0.01$.}
\label{2d:fig4}
\end{figure*}

Substituting the two-breather solution (\ref{2b1}) and (\ref{2b1a}) into (\ref{2d:a15}) we can obtain the general two-breather solution of (\ref{2d:eq1}).  In Figure~\ref{2d:fig4}, we display the evolution of two-breather solution of (\ref{2d:eq1}) for $F_1(t)=b_2 t^2+b_1 t+b_0$ with appropriate imaginary eigenvalues.  We investigate how the two-breather profiles evolve with $b_1=0.5$.  We obtain two AB solution from (\ref{2b1}) for the choice $\lambda_1=0.55i$ and $\lambda_2=0.75i$.  One AB developing with a time delay after another is shown in Figure~\ref{2d:fig4}(a).  This spatially periodic AB solution appear at two different times, say $t_1$ and $t_2$.  In Figure~\ref{2d:fig4}(b) we depict the case when one AB along with a RW for the choice $\lambda_1=0.55i$ and $\lambda_2=0.99i$.  We obtain two MB solution from (\ref{2b1}) by taking $\lambda_1=1.3 i$ and $\lambda_2=1.4 i$.  The evolution of two MBs with a spatial delay is shown in Figure \ref{2d:fig4}(c).  For $\lambda_1=0.5i$ and $\lambda_2=1.3i$ we observe that AB intersects the MB as shown in Figure~\ref{2d:fig4}(d).  When we increase the value of $b_1$, say for example from $0.5$ to $1.5$, the number of peaks increases in both the ABs which is demonstrated in Figure~\ref{2d:fig4}(e).  In the coexistence of AB and RW case, the number of peaks increases in ABs as shown in Figure~\ref{2d:fig4}(f).  The number of peaks increase in both the MBs in the constant density background as seen in Figure~\ref{2d:fig4}(g).  The intersection of AB-MB case, the number of peaks increase as depicted in Figure \ref{2d:fig4}(h).  It is clear from the above that when we tune the functional parameter $b_1$, in the obtained two-breather solution, the number of peaks and their densities increase in the considered region of $x-t$ plane.  
\subsection{Case 2}
Next we investigate the characteristics of RWs and breather profiles of (\ref{2d:eq1}) by considering another time-dependent functional parameter, namely $F_1(t)=1+\tanh(b_0t)$, with $b_0$ is a positive constant.  
\subsubsection{Characteristics of RWs}
Substituting this form into (\ref{2d:eq6}), we obtain the following localized solutions for (\ref{2d:eq1}), namely
\begin{subequations}
\label{2d:a17}
\begin{align}
\label{2d:a17b}
\psi(x,y,t) = & r_0\sqrt{b_0\sech^2{\Big(b_0t\Big)}}  \, U_j(X,T) \, \eta(x,y,t), \\
\label{2d:a17a}
\eta(x,y,t) = & \exp\left\{-i \left[\frac{1}{8k_0^2l_0^2 f}l_0^2+k_0\left(k_0 \right. \right. \right. \\  & \left. \left. \left.
+4l_0^2x\sqrt{b_0\sech{(b_0 t)}^2}+4k_0l_0y\sqrt{b_0\sech{(b_0 t)}^2}\right) \right. \right. \notag \\  & \left. \left. +(l_0^2+k_0^2(1-4b_0l_0^2(x^2+y^2)))\tanh{(b_0 t)}\right]\right\}. \notag 
\end{align}
\end{subequations}
where $U_j(X,T)$ are the RW solutions of the standard NLS equation. 
\begin{figure*}[!ht]
\begin{center}
\resizebox{0.99\textwidth}{!}{\includegraphics{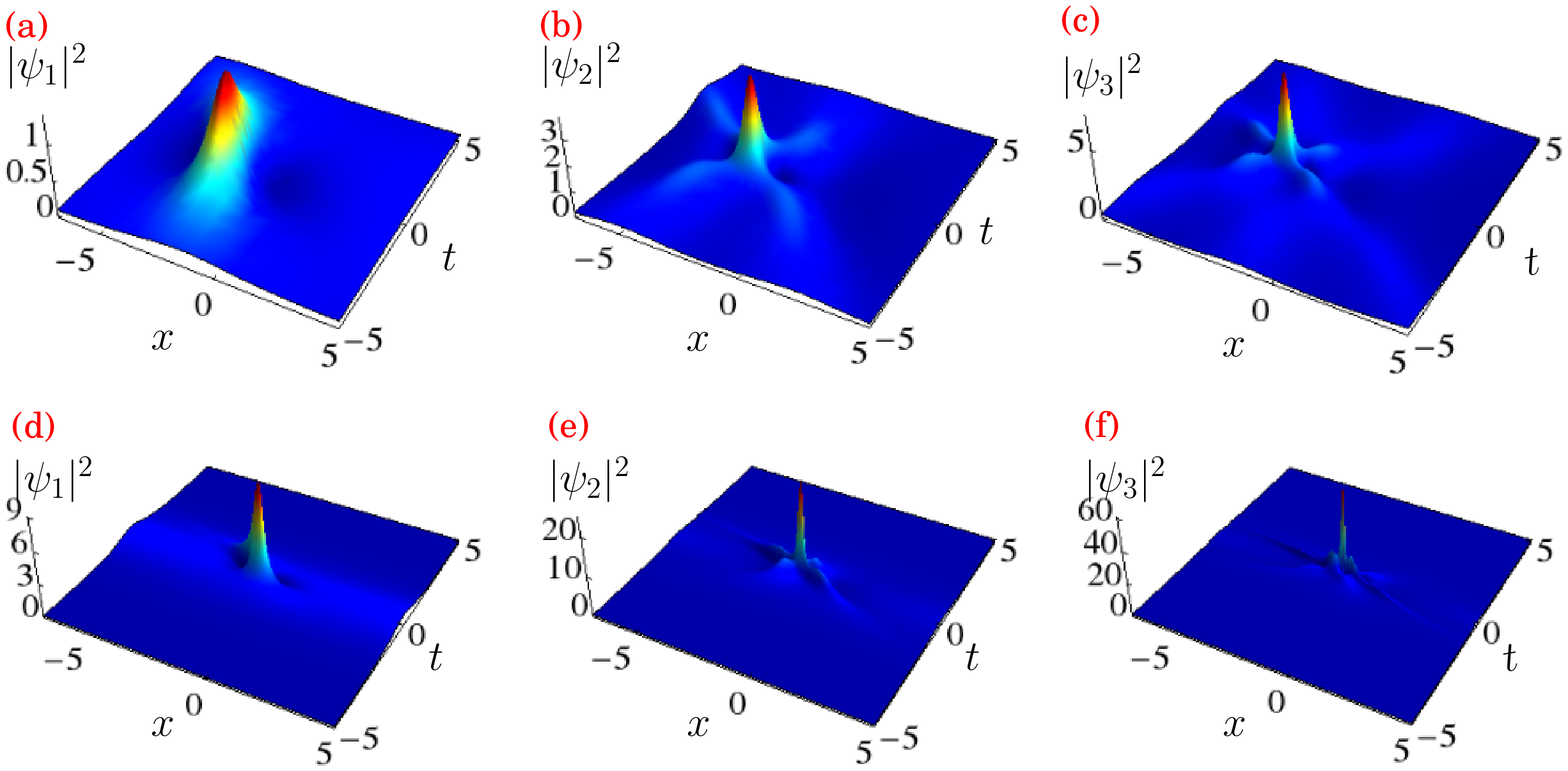}}
\end{center}
\caption{Density profiles. (a), (d) First-order RW, (b), (e) second-order RW, and (c), (f) third-order RW for $F_1(t)=1+\tanh(b_0t)$. The parameter $b_0$ is chosen as $0.15$ in (a)-(c) and $1.5$ in (d)-(f) .  The other parameters are fixed as $r_0=1.0$, $k_0=$ and $l_0=1$}
\label{2d:fig5}
\end{figure*}
In Figure \ref{2d:fig5}, we present the density profiles of first, second, and third-order RWs obtained from (\ref{2d:a17}) for (\ref{2d:eq1}). The density profile of the first-order RW for $b_0=0.15$ is given in Figure~\ref{2d:fig5}(a).  For a higher value of $b_0$, say $b_0=1.5$, we observe that the amplitude becomes higher and its density is more localized in time - see Figure \ref{2d:fig5}(d).  Figure~\ref{2d:fig5}(b) shows the density profile of second-order RW for $b_0=0.15$.  The second-order RW becomes more localized in time for a higher value of $b_0$ (=1.5) which is demonstrated in Figure~\ref{2d:fig5}(e).  A similar outcome has also been observed in third-order RW when we vary the parameter $b_0$.  Figure~\ref{2d:fig5}(c) shows the density profile of third-order RW with $b_0=0.15$.  The modified third-order RW with $b_0=1.5$ is shown in Figure~\ref{2d:fig5}(f).  From these figures it is clear that the underlying RWs become more localized in time and its density gets amplified when we increase the values of the constant in the functional parameter. 
\begin{figure*}[!ht]
\begin{center}
\resizebox{0.99\textwidth}{!}{\includegraphics{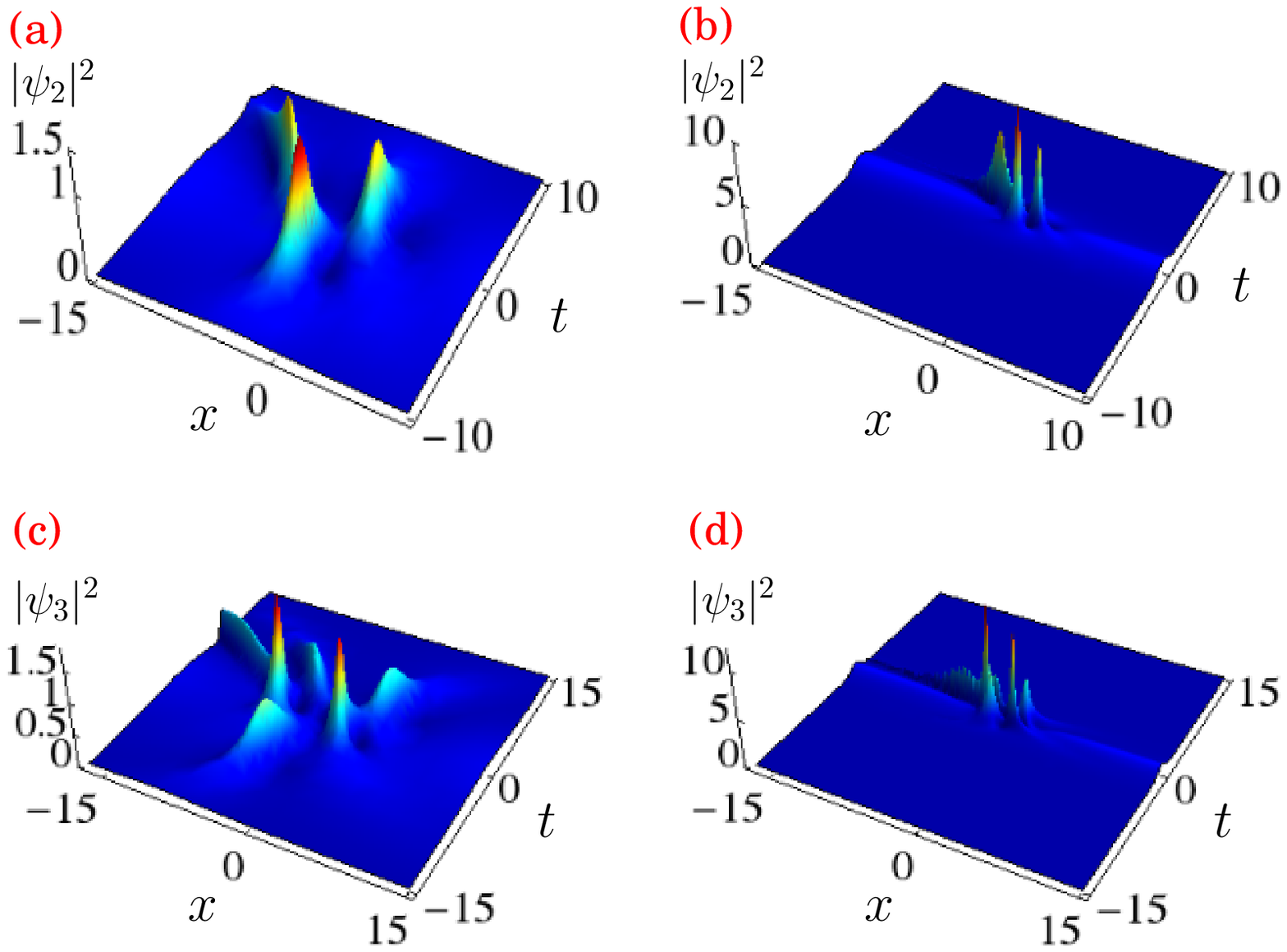}}
\end{center}
\caption{Density Profiles. (a), (b) Triplet RW, and (c), (d) sextet RW for $F_1(t)=1+\tanh(b_0t)$.  The parameter $b_0$ is chosen as $0.15$ in (a),(c) and $1.5$ in (b), (d).  The other parameters are fixed as $r_0=1.0$, $k_0=l_0=1$, $b_2=0.02$ and $b_0=0.01$.}
\label{2d:fig6}
\end{figure*}

Substituting the RW solution (\ref{trisoln}) into (\ref{2d:a17}), we obtain the triplet RW solution for (\ref{2d:eq1}). In Figures~\ref{2d:fig6}(a)-(b) we display the triplet pattern for $F_1(t)=1+\tanh(b_0t)$ with $s=1$ and $m=2$.  The triplet RW pattern for $b_0=0.15$ is shown in Figures~\ref{2d:fig6}(a).  For a higher value of $b_0$, say $b_0=1.5$, we observe that the triplet pattern becomes more localized in time and its amplitude has raised - see Figure~\ref{2d:fig6}(b).  We then move on to investigate how the structure of the third-order RW solution vary with respect to the free parameters $s_1,s_2,g$ and $h$.  The density profile of the sextet pattern is displayed in Figures~\ref{2d:fig6}(c)-(d) for the second functional parameter $F_1(t)=1+\tanh(b_0t)$ with $s_1=1$, $s_2=2$, $g=50$ and $h=50$.  When $b_0=0.15$ we obtain a set of six first-order RWs as shown in Figure~\ref{2d:fig6}(c).  For $b_0=1.5$, the sextet pattern becomes more localized in time with higher amplitude as shown in Figure~\ref{2d:fig6}(d). 
\begin{figure*}[!ht]
\begin{center}
\resizebox{0.99\textwidth}{!}{\includegraphics{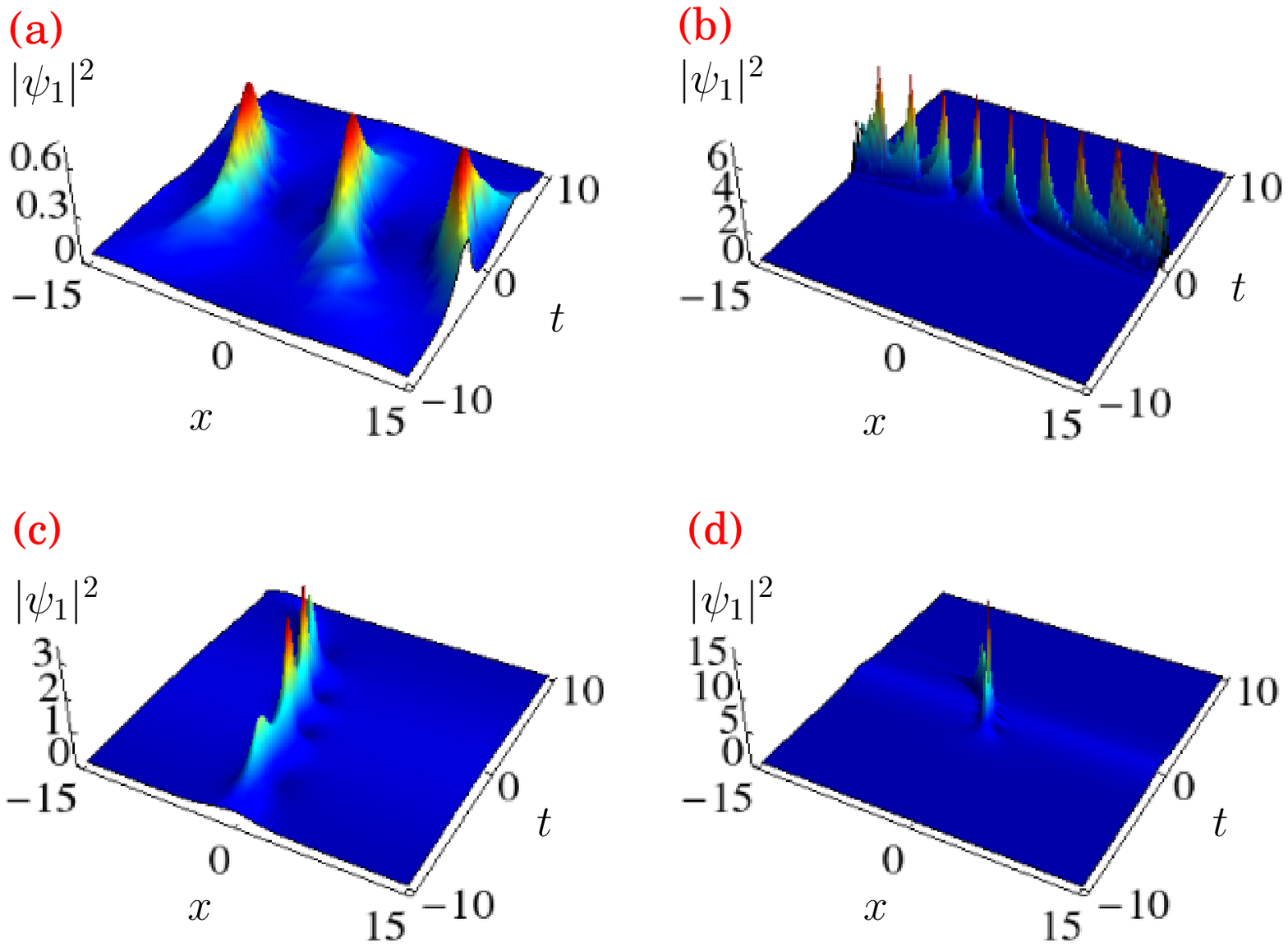}}
\end{center}
\caption{Density profiles. (a), (b) AB, and (c), (d) Ma breather for $F_1(t)=1+\tanh(b_0t)$.  The parameter $b_0$ is chosen as $0.15$ in (a),(c) and $1.5$ in (b), (d).  The other parameters are fixed as $r_0=1.0$, $k_0=1$, $l_0=1$, $b_2=0.02$ and $b_0=0.01$.}
\label{2d:fig7}
\end{figure*}

\subsubsection{Characteristics of breathers}
Substituting the expression (\ref{b1}) into (\ref{2d:a17}), we can get the breather solution of (\ref{2d:eq1}). Figure~\ref{2d:fig7} shows the density profile of breathers for the functional parameter $F_1(t)=1+\tanh(b_0t)$ with $b_0=0.15$.  The AB and MB arising from (\ref{2d:a17}) for the choice $\lambda=0.6 i$ and $1.95 i$ are displayed in Figures~\ref{2d:fig7}(a) and ~\ref{2d:fig7}(c), respectively.  When we increase the value of $b_0$, the number of peaks increases in the obtained breather profiles.  Unlike the earlier case here we notice that the underlying structures get stretched in space in the constant density background.  In the case of AB, when we increase the value of $b_0$ from $0.15$ to 1.5, the number of peaks increases from 3 to 9 which is demonstrated in Figure~\ref{2d:fig7}(b).  In the case of MB, the number of peaks is 3 when $b_0=0.15$ and the modified density profile becomes more localized in time for a higher value of $b_0$, say $b_0=1.5$, as illustrated in Figure~\ref{2d:fig7}(d).  The results reveal that when we tune the functional parameter $b_0$ in the obtained breather solution, the number of peaks and its amplitude increases in the AB profile whereas in the case of MB the modified density profiles become more localized in time and its amplitude becomes higher. 
\begin{figure*}[!ht]
\begin{center}
\resizebox{0.99\textwidth}{!}{\includegraphics{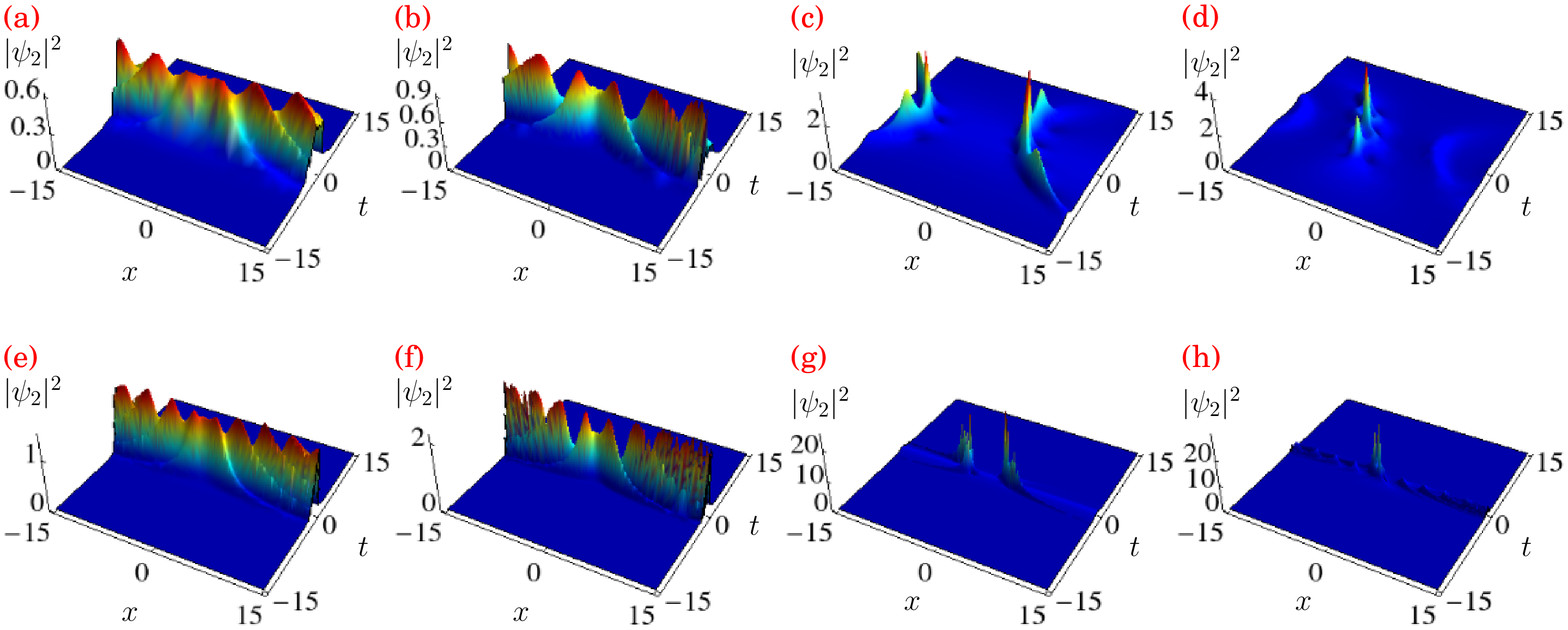}}
\end{center}
\caption{Modified density profiles.  (a), (e) Two AB, (b), (f) AB - RW, (c), (g) two Ma breather and (d), (h) the intersection of AB-MB for $F_1(t)=1+\tanh(b_0t)$. The parameter $b_0$ is chosen as $0.15$ in (a)-(d) and $1.5$ in (e)-(h).  The other parameters are fixed as $r_0=1.0$, $k_0=1$ and $l_0=1$.}
\label{2d:fig8}
\end{figure*} 

Substituting the two-breather solution (\ref{2b1}) into (\ref{2d:a17}) we obtain the general two-breather solution of (\ref{2d:eq1}).  In Figure~\ref{2d:fig8} we display the evolution of two-breather solution of (\ref{2d:eq1}) for $F_1(t)=1+\tanh(b_0t)$ with imaginary eigenvalues.  We analyze the two-breather profiles by setting the parameter $b_0=0.15$.  We can extract the ABs from (\ref{2d:a17}) by considering both the eigenvalues $\lambda_1$ and $\lambda_2$ are imaginary and less than one, that is $\lambda_1=0.55i$ and $\lambda_2=0.75i$.  The outcome is similar to the one observed earlier (Fig.~\ref{2d:fig4}(a)), that is one AB developing with a time delay after another.  However, here we obtain a collapsed state of two-breather profile as shown in Figure~\ref{2d:fig8}(a).  In Figure~\ref{2d:fig8}(b), we depict the case when one AB coexist with the RW for the choice $\lambda_1=0.55i$ and $\lambda_2=0.99i$.  When we restrict the eigenvalues as $\lambda_1=1.3i$ and $\lambda_2=1.4i$, we come across an evolution of bending profile of two MBs with spatial delay as shown in Figure \ref{2d:fig8}(c).  In the mixed case, that is $\lambda_1=0.5i$ and $\lambda_2=1.3i$, the AB intersects with MB as shown in Figure~\ref{2d:fig8}(d).  When we increase the value of $b_0$, we observe the following features: (i) both the ABs attain amplified structure as seen in Figure~\ref{2d:fig8}(e), (ii) in the coexistence of AB and RW case, the number of peaks increases in AB as shown in Figure~\ref{2d:fig8}(f), (iii) the density profiles of both the MBs become more localized in time and their densities have enhanced as seen in Figure~\ref{2d:fig8}(g) and (iv) in the intersection of AB-MB case, the number of peaks increases in the AB profile whereas MB gets more localized in time as demonstrated in Figure~\ref{2d:fig8}(h).  The above results reveal that the density of condensate atoms can be amplified by tuning the arbitrary parameter $b_0$ in the obtained two-breather solution of (\ref{2d:eq1}). We note here that one can also investigate the characteristics of localized solutions by considering negative values of $b_1$ and $b_0$.  However, when we choose the parameters $b_1$ and $b_0$ are negative we do not observe the typical characteristics of rogue waves and breathers in the solution profiles
\section{Vector localized solutions of quasi-two-dimensional two component BECs}
In this section, we consider quasi-two-dimensional two component BEC and construct the vector localized solutions.  The dynamics of a weakly interacting Bose gas at zero temperature is described by the following two coupled two-dimensional GP equation with time modulation \cite{pit:str}, that is
\begin{eqnarray}
i\psi_{1,t}+\frac{f}{2}\left(\psi_{1,xx}+\psi_{1,yy}\right)+R(t)\sum_{k=1}^{2} \vert \psi_k \vert^2 \psi_1 \nonumber \\ + \frac{1}{2}\beta^2(t)(x^2 +y^2) \psi_1=0, \nonumber \\
i\psi_{2,t}+\frac{f}{2}\left(\psi_{1,xx}+\psi_{2,yy}\right)+R(t)\sum_{k=1}^{2} \vert \psi_k \vert^2 \psi_2 \nonumber \\  + \frac{1}{2}\beta^2(t)(x^2 +y^2) \psi_2=0, 
\label{c2d:eq1}
\end{eqnarray}
where $\psi_j(x,t)$, $j=1,2$, are the macroscopic condensate wave functions, $t$ and $x$ are the temporal and spatial coordinates respectively, $f$ is the dispersion constant, $R(t)$ is the coefficient of nonlinearity and $\beta^2(t)$ is the axial trap frequency.  In constrast to the single-component BEC, the cross interactions between two components play a crucial role in the wave collision dynamics of two-component system.  Again we confine our attention only on attractive interatomic interactions.  

To investigate the dynamics of (\ref{c2d:eq1}), we transform the two-dimensional two coupled GP equation $(\ref{c2d:eq1})$ into two coupled NLS equation through the similarity reduction, 
\begin{equation}
\psi_j(x,y,t)=r(t)U_j(X,T)\exp[i \theta(x,y,t)], \;\; j=1,2,
\label{c2d:eq2}
\end{equation}
where $r(t)$ is the amplitude, $T(t)$ is the effective dimensionless time, $X(x,y,t)$ is the similarity variable and $\theta(x,y,t)$ is the phase factor which are all to be determined.  To determine these unknown functions we substitute (\ref{c2d:eq2}) into (\ref{c2d:eq1}) and obtain a set of PDEs.  Solving these PDEs, we find the same expressions for the functions $r(t)$, $X(x,y,t)$ and $\theta(x,y,t)$ as given in Eq. (\ref{2d:eq3}) except $T(t)$ which reads now $T(t)=\frac{1}{2}\int{R(t)r^2(t)}dt$.  Here we demand that the functions $U_1(X,T)$ and $U_2(X,T)$ should satisfy the following two coupled NLS equation, that is 
\begin{eqnarray}
\label{cnls}
i \frac{\partial U_1}{\partial T}+\frac{\partial ^2 U_1}{\partial X^2}+ 2 (|U_1|^2+|U_2|^2) U_1=0, \nonumber \\
i \frac{\partial U_2}{\partial T}+\frac{\partial ^2 U_2}{\partial X^2}+ 2 (|U_1|^2+|U_2|^2) U_2=0.
\end{eqnarray}
As our motivation is to investigate the characteristics of localized solutions of (\ref{c2d:eq1}) we consider the following form of solution \cite{fabio} to Eq. $(\ref{cnls})$, that is
\begin{eqnarray}
\label{csoln}
U_1(X,T)=\e^{2i\omega T}\left[\left(\frac{L}{B}\right)a_1+\left(\frac{M}{B}\right)a_2\right], \nonumber \\
U_2(X,T)=\e^{2i\omega T}\left[\left(\frac{L}{B}\right)a_2-\left(\frac{M}{B}\right)a_1\right],
\end{eqnarray}
where $L=\frac{3}{2}-8\omega^2 T^2-2a^2 X^2+8i\omega T+|p|^2 \e^{2aX}$, $M=4p(a X-2 i\omega T-\frac{1}{2})\e^{a X+i \omega T}$ and $B=\frac{1}{2}+8\omega^2 T^2+2a^2 X^2+|p|^2 \e^{2aX}$.  In the above, $a=\sqrt{a_1^2+a_2^2}$, $\omega=a^2$, $a_1$ and $a_2$ are arbitrary real parameters and $p$ is an arbitrary complex constant.  The solution (\ref{csoln}) is the semi-rational vector localized solution \cite{fabio}.  A main feature of this solution is that it has both exponential and rational dependence on coordinates.  Another interesting property of this solution is that the RWs coexist with dark-bright solitons when varying the complex arbitrary parameter $p$.  We will discuss these features in detail in the following sub-sections. The integrability condition (\ref{2d:eq4}) arises in the present analysis as well.  In other words, regardless the form of $F_1(t)$, as long as the integrability condition (\ref{2d:eq4}) is satisfied, we can obtain the solutions of $(\ref{c2d:eq1})$ in the form
\begin{subequations}
\label{c2d:eq6}
\begin{align}
\psi_j(x,t)= & r_0\sqrt{F_1'(t)}U_j(X,T) \eta(x,y,t), \\
\eta(x,y,t)= & \exp \left[i\left(-\frac{1}{2f}\left(\frac{k'(t)}{k(t)}x^2 + \frac{l'(t)}{l(t)}y^2\right) \right. \right. \notag \\  & \left. \left. -\frac{F_1'(t)}{2f}\left(\frac{x}{k(t)}+\frac{y}{l(t)}\right) -\frac{(k_0^2+l_0^2)F_1(t)}{8f k_0^2l_0^2}\right)\right], 
\end{align}
\end{subequations}
where $U_j(X,T)$, $j=1, 2$, are the solutions of the two coupled NLS Eq. (\ref{csoln}). Eq. (\ref{c2d:eq6}) confirms the occurrence of RW phenomena in BEC experiments and its potential applications.  
\subsection{Characteristics of vector rogue waves in two-dimensional BECs}
In this sub-section, we investigate the characteristics of vector RW solutions of two coupled GP Eq. $(\ref{c2d:eq1})$ by considering two different forms of functional parameter $F_1(t)$.
\begin{figure}[!ht]
\begin{center}
\resizebox{0.5\textwidth}{!}{\includegraphics{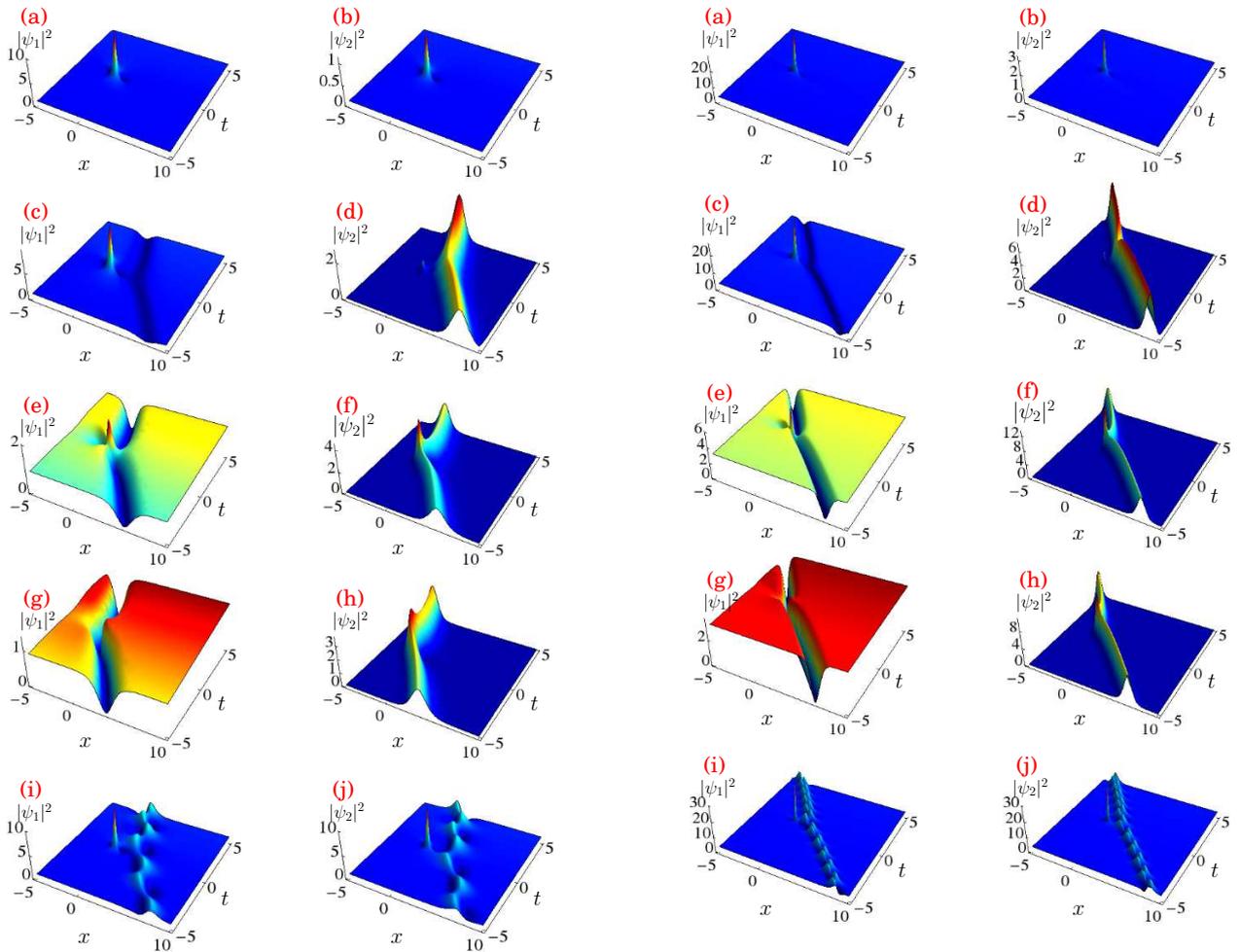}}
\end{center}
\caption{Density profiles. (a)-(b) Vector RWs for $p=0$. Dark-soliton with RW for (c) $p=0.2$, (e) $p=2.5$, (g) $p=11$. Bright-soliton with RW for (d) $p=0.2$, (f) $p=2.2$, (h) $p=12$. Panels (i)-(j): Breather-like wave with dark and bright contributions for $p=0.1i$. The other parameters are $F_1(t)=b_2 t^2+b_1 t+b_0$, $r_0=1.0$, $\beta_0=0.1$, $b=0.01$, and $\delta=0.01$.}
\label{cfig1}
\end{figure}
\begin{figure}[!ht]
\begin{center}
\resizebox{0.5\textwidth}{!}{\includegraphics{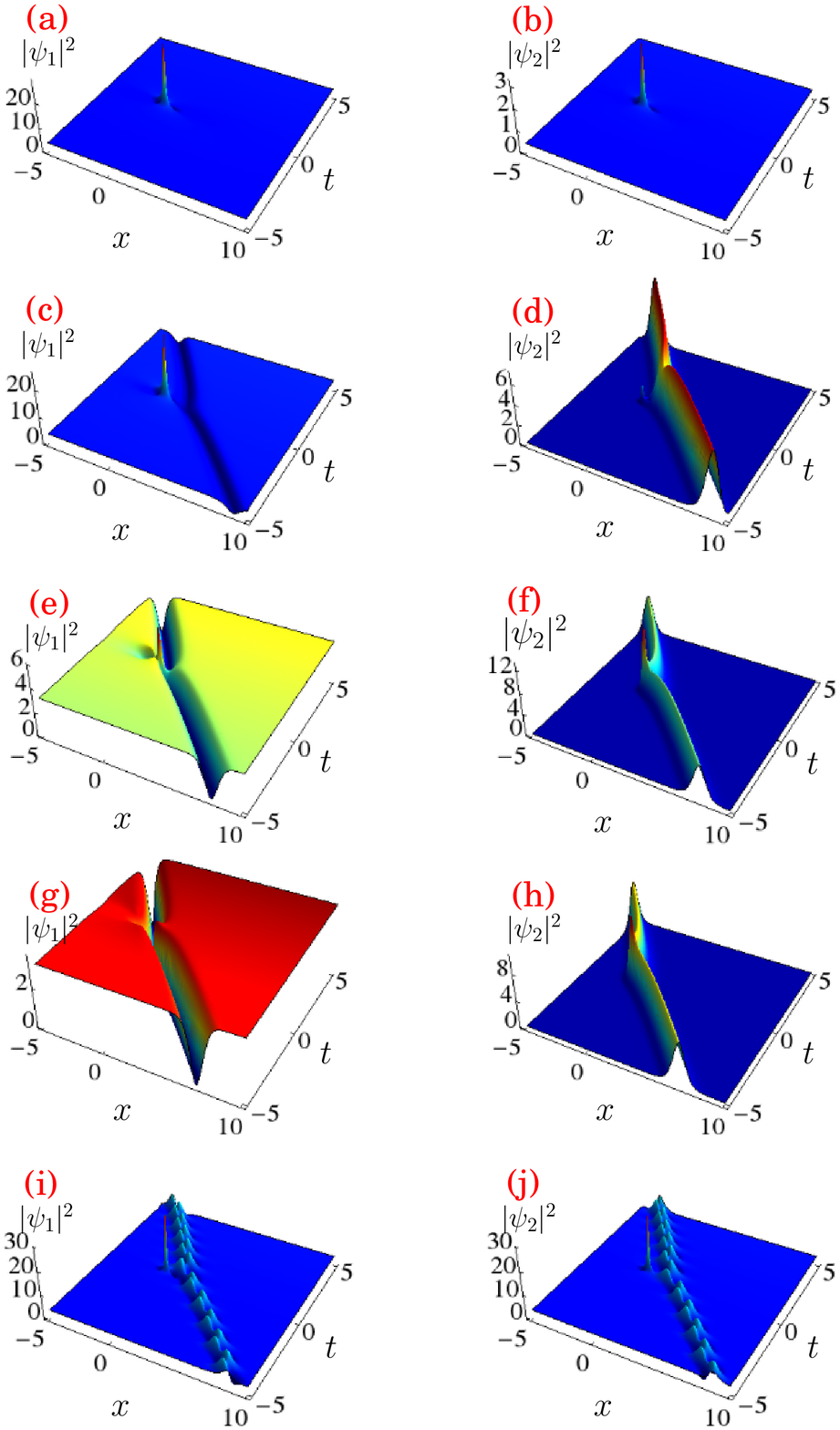}}
\end{center}
\caption{Modified density profiles.  All the descriptions are same as in Figure~\ref{cfig1} with $b_1=1.5$.}
\label{cfig2}
\end{figure}
\begin{figure}[!ht]
\begin{center}
\resizebox{0.5\textwidth}{!}{\includegraphics{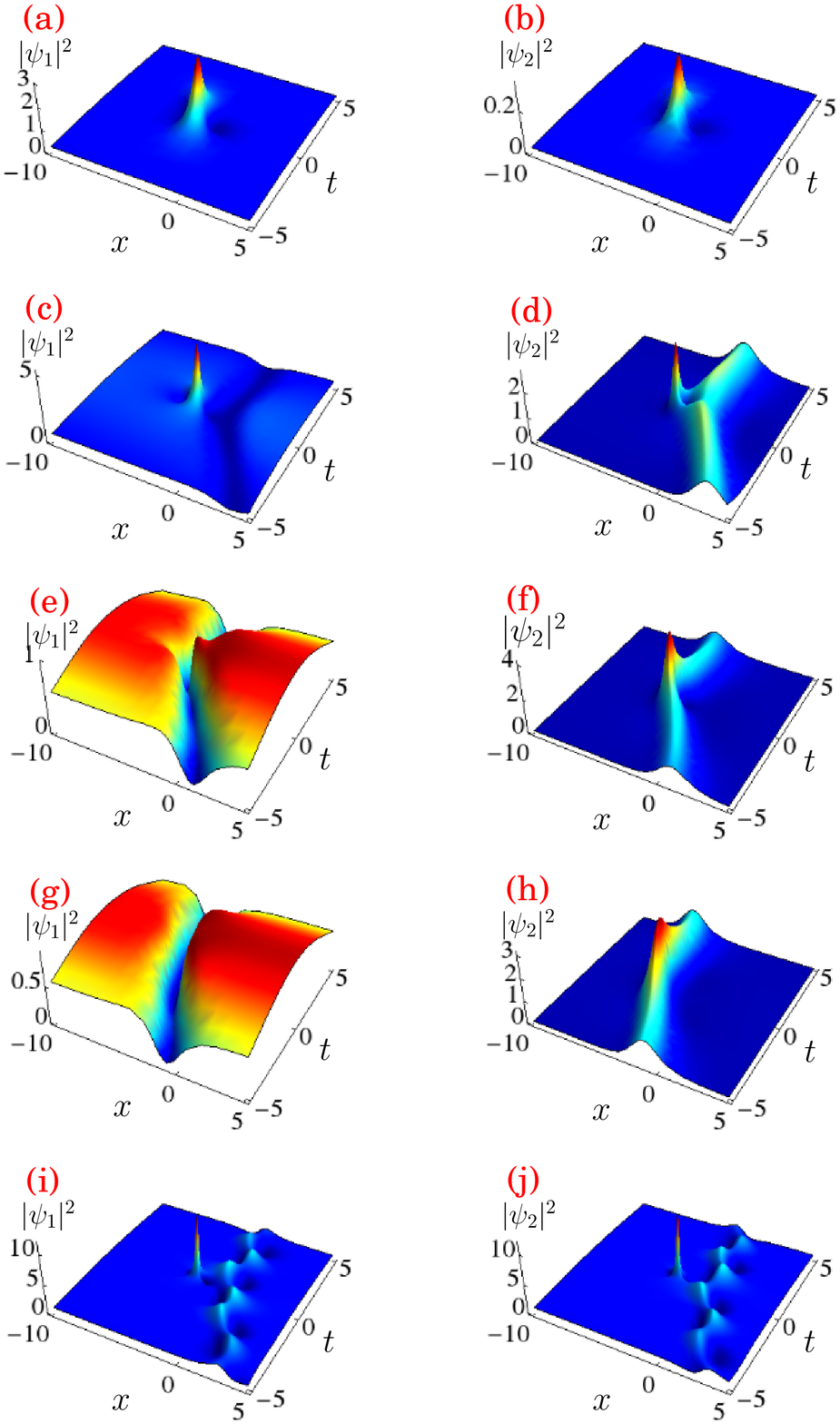}}
\end{center}
\caption{Density profiles. (a)-(b) Vector RWs for $p=0$. Dark-soliton with RW for (c) $p=0.2$, (e) $p=2.5$, (g) $p=11$. Bright-soliton with RW for (d) $p=0.2$, (f) $p=2.2$, (h) $p=12$. Panels (i)-(j): Breather-like wave with dark and bright contributions for $p=0.1i$. The other parameters are $F_1(t)=1+\tanh{b_0 t}$, $r_0=1.0$, $\beta_0=0.1$, $b=0.01$, and $\delta=0.01$.}
\label{cfig3}
\end{figure}
\begin{figure}[!ht]
\begin{center}
\resizebox{0.5\textwidth}{!}{\includegraphics{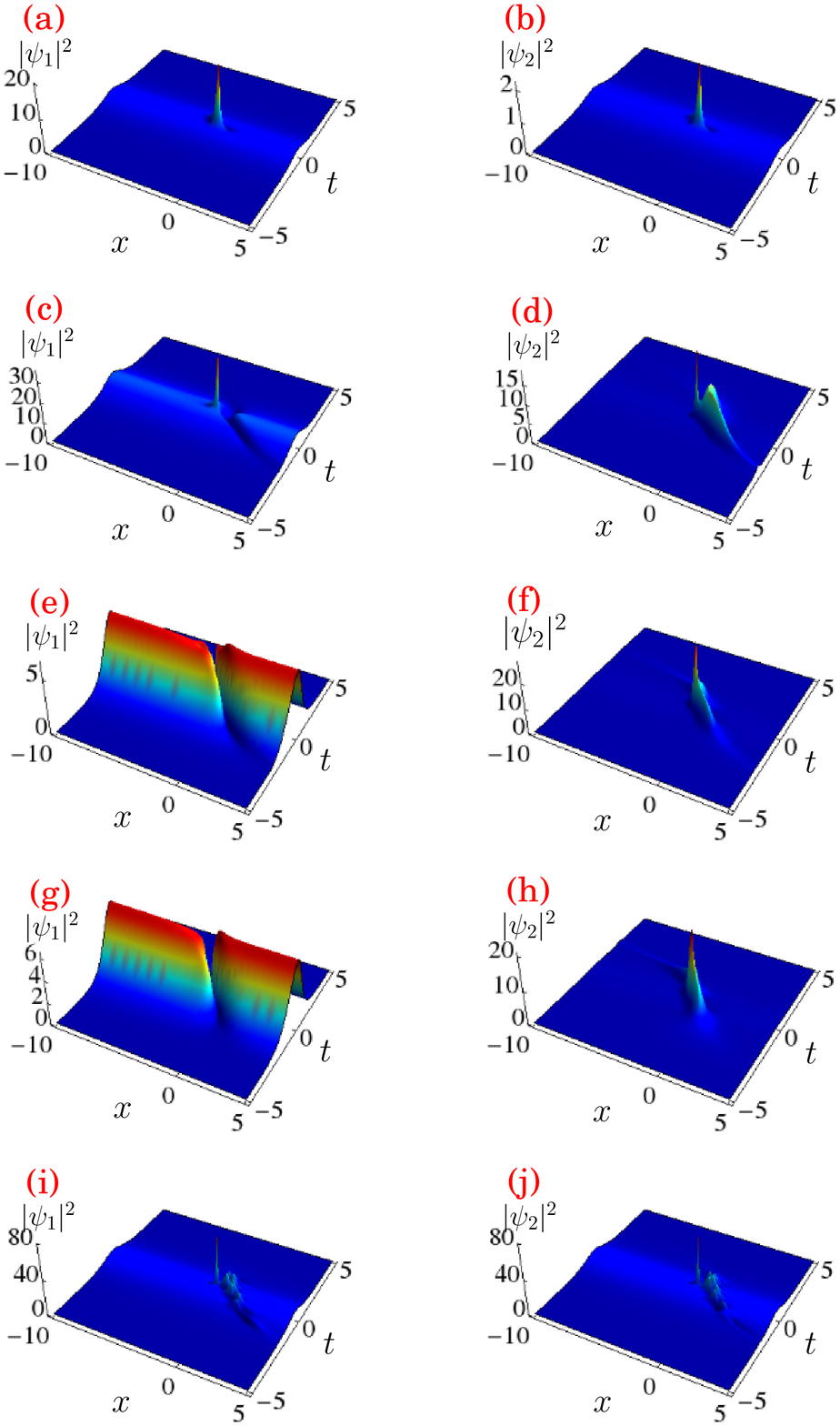}}
\end{center}
\caption{Modified density profiles. All the descriptions are same as in Figure~\ref{cfig3} with $b_1=1.5$.}
\label{cfig4}
\end{figure}
\subsubsection{Case 1} 
Substituting $F_1(t)=b_2 t^2+b_1 t+b_0 $ in the general solution (\ref{c2d:eq6}), we find 
\begin{align}
\label{c2d:a15}
\psi_j(x,y,t) = r_0\sqrt{2 b_2 t+b_1}\, U_j(X,T)  \eta(x,y,t),
\end{align}
where $\eta(x,y,t)$ is given in (\ref{2d:a15a}) and $U_j(X,T)$'s are the solutions of the coupled NLS equations (\ref{cnls}).  Unlike the earlier case, the semi-rational localized solutions $U_j(X,T)$ of the Manakov system (\ref{csoln}) generates certain additional features to the two component system Eq. (\ref{c2d:eq1}), as we see below.

In Figure~\ref{cfig1}, we analyze the qualitative nature of the density profiles of localized matter waves for the functional parameter $F_1(t)=b_2 t^2+b_1 t+b_0$.  To begin, we consider the case $p=0$ in Eq. (\ref{c2d:a15}) and obtain the vector RW solution, namely the Peregrine soliton.  As we see in Figures~\ref{cfig1}(a)-(b), the vector RWs arrive in a short time and decay to the background very quickly which in turn reveals the unstable nature of RWs. We note that the amplitude of first component is higher than the second component.

Next we consider the case $p\neq0$ in Eq. (\ref{c2d:a15}) and analyze the interaction of RW with a soliton which propagates with a nonconstant speed, for different values of $|p|$.  The interaction of RW with dark and bright solitons in $|\psi_1(x,t)|^2$ and $|\psi_2(x,t)|^2$ components is demonstrated in Figures~\ref{cfig1}(c)-(d), respectively by fixing $p=0.2$, $a_1=1.5$ and $a_2=0$.  When we increase the value of $p$ in the RW solution (\ref{c2d:a15}) we observe that the RW merges with the dark- and bright solitons as illustrated in Figures~\ref{cfig1}(e)-(h). After the merging the amplitude of first component has suppressed whereas the amplitude of second component has enhanced.  When $p=2.5$ the RW completely merged with dark- and bright-solitons as seen in Figures~\ref{cfig1}(e)-(f).  At $p=11$, the RW cannot be identified separately while the resulting dark-bright distribution in both the components appear as a boomeron-type soliton as shown in Figures~\ref{cfig1}(g)-(h).  From these observations we conclude that for small values of $|p|$, the RW and dark-bright solitons keep away from each other and for higher values of $|p|$, the RW and dark-bright solitons merge with each other.  These facts can be observed in each one of the components. 

In the above, we have considered only the real values of $p$.  We can also consider the complex values for $p$.  In the later case the solution (\ref{c2d:a15}) tells us that the matter wave behave like a breather.  The RW interacts with a breather-like wave resulting from the interference between the dark and bright contributions is shown in Figures~\ref{cfig1}(i)-(j) for $p=0.1i$. The interaction of RWs with dark and bright distributions in each component represent the exchange of condensate atoms between RW and soliton/breather profiles.  

Figure~\ref{cfig2} shows the modified density profiles for the same set of parameters, constituting Figure~\ref{cfig1}, but for $b_0=1.5$.  The amplitude of RWs in each component gets sharpened as shown in Figures~\ref{cfig2}(a)-(b).  In Figures~\ref{cfig2}(c)-(d), we present the interaction of RW with the dark and bright solitons which in turn reveals that there is an exchange of atoms between RW and the soliton and they become more localized in the constant density background.  More localization of condensate atoms in the constant density background can be seen in Figures~\ref{cfig2}(e)-(h).  The modified RW with breather-like wave in each one of the components can be seen in Figures~\ref{cfig2}(i)-(j). In these figures the number of peaks has increased in the breather-like profiles.  Comparing the Figures~\ref{cfig1} and \ref{cfig2}, we conclude that for a higher value of $b_1$, the density profile of condensate atoms become more steepened in the constant density background.
\subsubsection{Case 2}
Now we investigate how the solution profiles are influenced for the functional parameter $F_1(t)=1+\tanh(b_0t)$.  Plugging this expression into (\ref{c2d:eq6}), we obtain the localized solutions of (\ref{c2d:eq1}) in the form,
\begin{align}
\label{c2d:a22}
\psi_j(x,y,t) = r_0\sqrt{b_0\sech\Big(b_0t\Big)^2}  \, U_j(X,T) \, \eta(x,y,t), \;\; j=1,2
\end{align}%
where $\eta(x,y,t)$ is given in (\ref{2d:a17a}) and $U_j(X,T)$'s are the solutions of the coupled NLS equations (\ref{csoln}).  The qualitative nature of various density profiles of (\ref{c2d:eq1}) for $F_1(t)=1+\tanh(b_0t)$ are presented in Figure~\ref{cfig3} with $b_0=0.15$.  The outcome is similar to the previous case except that the densities are now higher than the previous one. So we do not discuss this case separately.  We choose $b_0=1.0$ and fix all other parameters are exactly the same as given in Figure~\ref{cfig3}.  We observe that the vector RWs exist on a higher density background now - see Figures~\ref{cfig4}(a)-(b).  In Figures~\ref{cfig4}(c) and \ref{cfig4}(d), we display the interaction of RW with the bending profile of dark and bright solitons in the first and second components, respectively.  The RW together with dark-soliton is more localized in time and delocalized in space in the $|\psi_1(x,t)|^2$ component as seen in Figures~\ref{cfig4}(e) and \ref{cfig4}(g). The RW together with bright-soliton becomes more localized/sharpened as shown in Figures~\ref{cfig4}(f) and \ref{cfig4}(h).  In these figures, we notice that the characteristic feature of RWs (troughs) is completely disappeared and the dark-soliton feature dominates in the first component whereas in the second component bright-soliton disappeared and the characteristic feature of RW dominates.  Figures~\ref{cfig4}(i)-(j) we show the modified structure of the RW with breather-like waves.  From these outcome we conclude that the density fluctuations of condensate atoms are more localized in time and delocalized in space, but their amplitudes have increased (RWs alongwith soliton) in both the components.
\section{Conclusion}
We have constructed RW and breather solutions for the quasi two-dimensional GP equation with time-dependent interatomic interaction and external trap through similarity reduction technique. We have derived these solutions by transforming the GP equation into constant coefficient NLS equation.  We have investigated the characteristics of the constructed RW solutions in detail for two different forms of the functional parameter.  We have demonstrated that the amplitude of RWs increases when a parameter appearing in the functional parameter is varied.  We have also considered the generalized RW solutions of (\ref{2d:eq1}) and brought out the triplet and sextet patterns exhibited by (\ref{2d:eq1}).  We have probed how these periodic localized waves change in the constant density background when we tune the parameter in the obtained breather solutions.  In addition to the above, we have constructed the vector localized solutions for the coupled quasi-two-dimensional BECs and investigated the characteristics of the localized density profiles by varying the functional parameter.  When we increase the value of the first chosen functional parameter we have noticed that the density profiles become more localized and their amplitudes get sharpened in the constant density background.  As far as the second functional parameter is concerned the underlying structures become more localized in time and delocalized in space.  Our results may provide possibilities to manipulate RWs experimentally in the two-dimensional BEC system.

\begin{acknowledgement}
KM thanks the University Grants Commission (UGC-RFSMS), Government of India, for providing a research fellowship. The work of MS forms part of a research project sponsored by NBHM, Government of India. 
\end{acknowledgement}
\begin{appendix}
\section{}
Several localized and periodic solutions of the standard NLS equation are reported in the literature \cite{akmv:anki,eleon,kadz}.  Equation $(\ref{nls})$ admits $N^{th}$ order RW solution which can be written in the following form \cite{akmv:anki},
\begin{align}
\label{a6}
U_j(X,T)=\left[(-1)^j+\frac{G_j(X,T)+iTH_j(X,T)}{D_j(X,T)}\right]\exp{(iT)},
\end{align}
where $j=1,2,...,G_j,H_j$ and $D_j$ are polynomials in the variables $X$ and $T$.

For the first-order $(j=1)$ RW solution $G_1=4$, $H_1=8$ and $D_1=1+4X^2+4T^2$ so that we get \\ $U_1=\left(4\frac{1+2iT}{1+4X^2+4T^2}-1\right)\exp{(iT)}$.  For convenience we multiply this expression by $-1 = \exp[i\pi]$ and consider the solution in the following form, that is
\begin{align}
\label{a8}
U_1(X,T)=\left(1-4\frac{1+2iT}{1+4X^2+4T^2}\right)\exp{[iT]}.
\end{align}
For the second-order $(j=2)$ RW solution, the functions $G_2$, $H_2$, and $D_2$ read \cite{akmv:anki}
\begin{align}
\label{a9}
G_2 = & \frac{3}{8}-3X^2-2X^4-9T^2-10T^4-12X^2T^2, \notag \\
H_2 = & \frac{15}{4}+6X^2-4X^4-2T^2-4T^4-8X^2T^2
\end{align}
and
\begin{align}
D_2 = &  \frac{1}{8}\left(\frac{3}{4}+9X^2+4X^4+\frac{16}{3}X^6+33T^2+36T^4 \right. \notag \\
& \left.+\frac{16}{3}T^6-24X^2T^2+16X^4T^2+16X^2T^4\right). \notag
\end{align}
With these expressions, Eq. (\ref{a6}) now becomes
\begin{align}
\label{a11}
U_2(X,T)=\left[1+\frac{G_2+iTH_2}{D_2}\right]\exp{(iT)}.
\end{align}
For the third-order $(j=3)$ RW solution, we have
\begin{align}
\label{a12}
U_3(X,T)=\left[-1+\frac{G_3+iTH_3}{D_3}\right]\exp{(iT)},
\end{align}
where
\begin{align}
G_3(X,T)= & g_0 + (2 T)^2 g_2 + (2 T)^4 g_4 + (2 T)^6 g_6  + (2 T)^8 g_8 \nonumber \\ & + (2 T)^{10} g_{10},
\end{align}
with
\begin{align}
\label{solg}
g_0 = &  1 - (2 X)^2 - \frac{2}{3} (2 X)^4 + \frac{14}{45} (2 X)^6 + \frac{(2 X)^8}{45} + \frac{(2 X)^{10}}{675},\notag \\
g_2 = &  -3 - 20 (2 X)^2 + \frac{2}{3}(2 X)^4 - \frac{4}{45} (2 X)^6 + \frac{(2 X)^8}{45},\notag \\
g_4 = &  2 \left[-\frac{17}{3} + 5 (2 X)^2 - \frac{(2 X)^4}{3^2} + \frac{(2 X)^6}{3^3}\right]\notag \\
g_6 = &  \frac{2}{45}\left[73 +14 (2 X)^2 + \frac{7}{3}(2 X)^4\right],\notag \\
g_8 = &  \frac{1}{15} (11 + (2 X)^2), \;\;\;\; 
g_{10} = \frac{11}{675}
\end{align}
and 
\begin{align}
H_3(X,T)= & h_0 + (2 T)^2 h_2 + (2 T)^4 h_4 + (2 T)^6 h_6  + (2 T)^8 h_8 \notag \\ & + (2 T)^{10} h_{10}, 
\end{align}
with
\begin{align}
h_0 = & 2 \left[7 + 7 (2 X)^2 - 2 (2 X)^4 - \frac{2}{3^2} (2 X)^6 - \frac{(2 X)^8}{45} \right. \notag \\ & \left.  + \frac{(2 X)^{10}}{675} \right], \notag \\
h_2 = &  \frac{2}{3} \left[-11 - 28 (2 X)^2 - 2 (2 X)^4 - \frac{28}{45} (2 X)^6 + \frac{(2 X)^8}{45}\right],\notag \\
h_4 = &  \frac{4}{15} \left[-107 + 19 (2 X)^2 - \frac{7}{3} (2 X)^4 + \frac{(2 X)^6}{3^2}\right],\notag \\
h_6 = &  \frac{4}{45} \left[-29 - 2 (2 X)^2 +\frac{(2 X)^4}{3}\right],\notag  \\
h_8 = & \frac{2}{3^3} \left[1 + \frac{(2 X)^2}{5}\right], \;\;\; h_{10} =\frac{2}{675}.
\end{align}
The denominator is represented by
\begin{align}
D_3(X,T)=& d_0 + (2 T)^2 d_2 + (2 T)^4 d_4 + (2 T)^6 d_6 + (2 T)^8 d_8 \notag \\ & + (2 T)^{10} d_{10}+ (2 T)^{12} d_{12},\notag 
\end{align}
where
\begin{align}
d_0 = & \frac{1}{2^3} \left[1 + 6 (2 X)^2 + \frac{5}{3} (2 X)^4 + \frac{52}{45} (2 X)^6 + \frac{(2 X)^8}{15}  \right. \notag \\ & \left. + \frac{2}{675} (2 X)^{10} + \frac{(2 X)^{12}}{2025}\right],\notag \\
d_2 = &  23 - 9 (2 X)^2 + \frac{10}{3} (2 X)^4 + \frac{2}{15} (2 X)^6 - \frac{(2 X)^8}{45} \notag \\ & + \frac{(2 X)^{10}}{675},\notag \\
d_4 = &  2 \left[71 + \frac{116}{3} (2 X)^2 - \frac{2}{3} (2 X)^4 - \frac{4}{45} (2 X)^6 + \frac{(2 X)^8}{135}\right],\notag \\
d_6 = &  \frac{32}{3} \left[\frac{17}{3} + 5 (2 X)^2 +\frac{(2 X)^4}{45} + \frac{(2 X)^6}{135}\right],\notag \\
d_8 = &  \frac{32}{15} \left[\frac{83}{3} + 2 (2 X)^2 +\frac{(2 X)^4}{3^2}\right],\notag \\
d_{10} = &  \frac{2^8}{225} \left[7 + \frac{(2 X)^2}{3}\right], \;\;\; d_{12} = \frac{2^9}{2025}.
\end{align}
To derive the RW solutions of (\ref{2d:eq1}) we recall these expressions. 
\end{appendix}

\end{document}